\documentclass[pdflatex,sn-mathphys-num]{sn-jnl}

\usepackage{graphicx}%
\usepackage{multirow}%
\usepackage{amsmath,amssymb,amsfonts}%
\usepackage{amsthm}%

\usepackage{mathrsfs}%
\usepackage[scaled=1.0]{rsfso}  

\usepackage[title]{appendix}%
\usepackage{xcolor}%
\usepackage{textcomp}%
\usepackage{manyfoot}%
\usepackage{booktabs}%
\usepackage{algorithm}%
\usepackage{algorithmicx}%
\usepackage{algpseudocode}%
\usepackage{listings}%
\usepackage{ulem}

\usepackage{lineno}

\theoremstyle{thmstyleone}%
\theoremstyle{thmstyletwo}%

\theoremstyle{thmstylethree}%

\title[Article Title]{Resonate-and-Fire Photonic-Electronic Spiking Neurons 
for Fast and Efficient Light-Enabled Neuromorphic Processing Systems}

\author*[1]{\fnm{Andrew} \sur{Adair}}\email{andrew.adair@strath.ac.uk}
\author[1]{\fnm{Dafydd} \sur{Owen-Newns}}
\author[1]{\fnm{Giovanni} \sur{Donati}}
\author[1]{\fnm{Joshua} \sur{Robertson}}
\author[2]{\fnm{Jos\'e} \sur{Figueiredo}}
\author[3]{\fnm{Eduard} \sur{Wasige}}
\author[3]{\fnm{Qusay} \sur{Al-Taai}}
\author[4]{\fnm{Bruno} \sur{Romeira}}
\author[5]{\fnm{Mat\v{e}j} \sur{Hejda}}
\author[1]{\fnm{Antonio} \sur{Hurtado}}

\affil[1]{\orgdiv{Institute of Photonics, SUPA Dept. of Physics, University of Strathclyde, Glasgow, United Kingdom}}
\affil[2]{\orgdiv{LIP and Departamento de F\'isica de Faculdade de Ci\^encias de Universidade de Lisboa, Lisbon, Portugal}}
\affil[3]{\orgdiv{High Frequency Electronics Group, University of Glasgow, Kelvin Building, Glasgow, United Kingdom}}
\affil[4]{\orgdiv{International Iberian Nanotechnology Lab., Braga, Portugal}}
\affil[5]{\orgdiv{Large-Scale Integrated Photonics Lab, Hewlett Packard Enterprise Labs, Diegem, Belgium}}

\begin{document}


\abstract{
Neuromorphic computing seeks to replicate the spiking dynamics of biological neurons 
for brain-inspired computation. 
While electronic implementations of artificial spiking neurons have dominated to date, 
photonic approaches are attracting increasing research interest 
as they promise ultrafast, energy-efficient operation with low-crosstalk and high bandwidth. 
Nevertheless, existing photonic neurons largely mimic integrate-and-fire models, 
but neuroscience shows that neurons also encode information through richer mechanisms, 
such as the frequency and temporal patterns of spikes. 
Here, we present a photonic–electronic resonate-and-fire (R\&F) spiking neuron 
that responds to the temporal structure of high-speed optical inputs. 
This is based on a light-sensitive resonant tunnelling diode that produces excitable spikes 
in response to nanosecond, low-power ($<100$\,\textmu W) optical signals 
at infrared telecom wavelengths. 
We experimentally demonstrate control of R\&F dynamics 
through inter-pulse timing of the optical stimuli and applied bias voltage, 
achieving bandpass filtering of both analogue and digital inputs. 
The R\&F neuron also supports optical fan-in via wavelength-division multiplexed inputs 
from four vertical-cavity surface-emitting lasers (VCSELs). 
This electronic–photonic neuron exhibits key functionalities --- including 
spike-frequency filtering, temporal pattern recognition, and digital-to-spiking conversion
--- critical for neuromorphic optical processing. 
Our approach establishes a pathway toward low-power, high-speed temporal information processing 
for light-enabled neuromorphic computing.
}



\maketitle

\section{Introduction}\label{sec1}

As the popularity of artificial intelligence (AI) continues to grow exponentially, 
so does the demand for efficient and fast (state-of-the-art) computing hardware. 
Novel neuromorphic (brain-inspired) processing systems, 
drawing direct inspiration from the computational capabilities of the brain 
and biological networks of neurons are attracting increasing research interest. 
Traditionally, neuromorphic computing research has mainly focused in electronic technology implementations, with developments such as the Loihi and SpiNNaker platforms 
by Intel and the University of Manchester, among others
\cite{Davies2018} \cite{Furber2013} \cite{DeBole2019}.
Nevertheless, photonic approaches for neuromorphic systems are receiving growing interest, 
given their potential to improve processing capability and efficiency 
by leveraging the low operating powers and inherent high-speed and bandwidths of light-based technologies. 
In recent years, approaches based on different photonic platforms, 
such as semiconductor lasers \cite{Robertson2020A} \cite{Robertson2020B} \cite{Bueno2021}, 
micro-ring resonators (MRRs) \cite{Donati2024}, 
and semiconductor optical amplifiers (SOAs) 
\cite{Vandoorne2014} \cite{Toole2016} \cite{Kravtsov2011}, 
to name but a few, have been proposed for implementation of artificial optical neuronal models and neural networks.

\begin{figure*}[t!]
\centering
\includegraphics[width=1\linewidth]{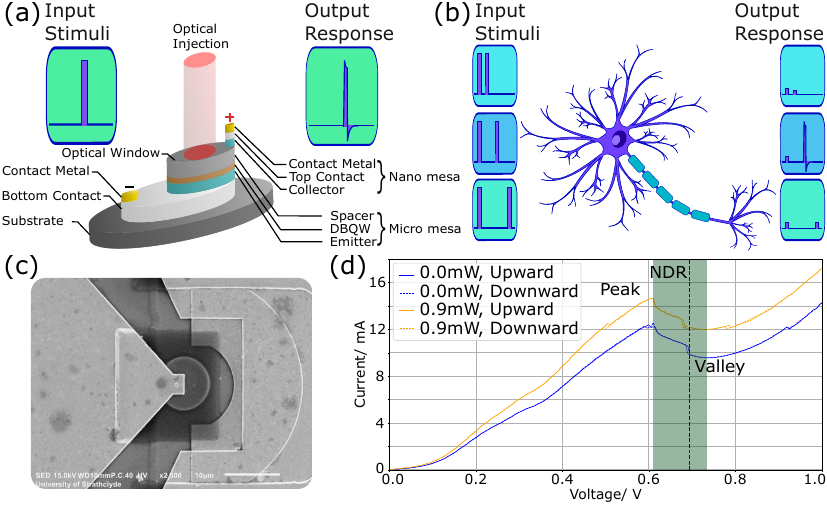}
\caption{\label{fig:Fig1___Intro}
    a) Optically-sensitive RTD neuron 
    displaying excitable spike firing in response to input light stimuli.
    b) Schematic diagram of a biological neuron 
    exhibiting resonate-and-fire spiking responses.
    c) SEM image of the light-sensitive RTD of this work. 
    d) Experimentally-measured I-V curve of the RTD neuron 
    when operated in dark conditions ($0$ mW) 
    and under infrared illumination of CW light at $1550$ nm with optical power of $0.9$ mW.
    The NDR region of the RTD is shaded in green 
    and the black dashed line marks a typical biasing point in the valley region.
    }
\end{figure*}

Another platform that has also recently shown great promise 
in realising light-enabled neuronal functionalities for neuromorphic photonic processing systems
is the resonant tunnelling diode (RTD) \cite{Ortega2022}. 
An RTD is a mesoscopic semiconductor device that features a double-barrier quantum well (DBQW) 
(as seen in Fig.\ref{fig:Fig1___Intro}a) in its structure. 
This configuration enables quantum tunnelling effects 
that lead to a highly nonlinear current-voltage (I-V) relationship, 
defining a region of negative differential resistance (NDR), 
around which neuronal-like nonlinear dynamical regimes and excitability can be achieved \cite{Romeira2013A}. 
Moreover, the quantum tunnelling effects at the core of the operation of RTDs as artificial neuronal models, are produced at extremely high bandwidths 
(e.g. these have been used for THz signal generation) 
\cite{Cimbri2022} \cite{Nishida2019}; 
thus, offering promise for ultrafast and energy-efficient operation. 
Furthermore, RTDs can be made optically sensitive 
by adding photo-absorption layers in their epi-layer stack structure 
to allow for operation under direct injection of optical signals \cite{Zhang2021}. 
As a result, recent works have reported experimentally and in theory 
on the neuronal excitability properties of RTDs under electronic and photonic excitation 
\cite{AlTaai2023}, 
including all-or-nothing spike firing regimes, thresholding and refractoriness 
\cite{Hejda2022B}. 
Moreover, RTDs have also recently demonstrated spike rate coding \cite{Donati2025} 
and flip-flop memory capabilities \cite{Donati2024B}; 
therefore strengthening their position as exciting candidates 
for artificial photonic-electronic spiking neurons 
towards use in future neuromorphic photonic computing systems. 
Additionally, integrate-and-fire spiking regimes have been described theoretically 
and more recently also experimentally in RTD neurons, 
demonstrating the feasibility of RTDs 
as photonic-electronic Leaky-Integrate-and-Fire (LIF) neuronal models 
\cite{Hejda2022A} \cite{Robertson2024}.

However, there exists another important class of biological neurons, termed resonator neurons, 
whose behaviour cannot be described by the integrator model \cite{Izhikevich2000}. 
This resonate-and-fire (R\&F) computational neuronal model \cite{Izhikevich2001} 
has gained traction due to its mathematical simplicity and biological plausibility 
\cite{Huber2025}. 
Resonator-type biological neurons are suspected to play important roles in brain function 
for timing in motor coordination, as well as specifying neural connectivity during development 
\cite{Llinas1988}. 
Yet, apart from isolated reports 
(for a review on the implementation of R\&F in CMOS, see \cite{Le2023}, 
and for a resonator spiking neuron model simulated with a silicon microring 
using a modulator and photodiode scheme, see \cite{Tamura2024}), 
their implementation and realisation with photonic hardware remains under-researched.

The R\&F effect behind resonator spiking neurons can be best explained 
by considering the neuron’s behaviour (change in state variable) 
following a sub-threshold input perturbation \cite{Izhikevich2000}. 
During relaxation from an impulse, 
the neuron’s voltage oscillates around its rest voltage before converging. 
Hence, after an impulse, 
the threshold required to trigger spike firing in the system changes with time. 
If a pair of sub-threshold input stimuli 
(each without sufficient intensity to elicit a spike on their own) arrives at the neuron, 
then the response is dependent on the precise timing of the secondary stimulus 
relative to the phase of the relaxation oscillation induced by the primary stimulus. 
This sensitivity of the neuron to the temporal separation of incoming sub-threshold stimuli doublets 
gives rise to bandpass filtering capabilities 
(as seen in Fig.\ref{fig:Fig1___Intro}b), 
resulting in spike firing events in the system achieved 
when the inter-(sub-threshold-)stimulus time matches the resonance period of the neuron. 
Additionally, inhibition of a supra-threshold stimulus can also occur 
when an input (that would otherwise trigger a spike) enters the system 
at a time when the neuron’s voltage is in a ‘trough’ of a relaxation oscillation 
triggered by a previous sub-threshold input, such that no spike is fired. 
Interestingly, another characteristic of R\&F neurons 
is that a train of input pulses with reverse-polarity 
(that is, shifting the system away from its spike firing threshold)
arriving synchronous with the neuron resonance frequency 
will cause the oscillation amplitude in the system to build until spiking threshold is reached 
\cite{Izhikevich2001}.
These neural functionalities of spike-firing under sub-threshold in-resonance inputs, 
excitability from reverse-polarity stimuli, and inhibition of supra-threshold pulses 
achieved in the R\&F neuronal system are not observed in the LIF model. 
This uncovers additional computational capabilities for processing data in the frequency domain, 
as well as allowing new methods to encode data for neuromorphic event-based computing tasks. 
As a result, neuromorphic spiking neural network models based upon R\&F neurons 
have been demonstrated as enabling various practical applications, including 
visual odometry for tracking location and rotation \cite{Renner2024B}, 
in machine vision for visual scene understanding \cite{Renner2024A}, 
and in associative memory models for image retrieval \cite{Paxon2019}.

In this work, we demonstrate experimentally, for the first time, 
a photonic-electronic R\&F spiking neuron.
Our approach is based on a photo-detecting RTD device 
capable of firing excitable electrical spikes 
in response to fast and low-power optical input signals. 
Furthermore, our approach is verified  theoretically using an analytical simulation model, 
yielding excellent agreement with the experimental findings (see supporting information).
We show the successful achievement of resonate-and-fire effects in RTD neurons 
when subjected to externally injected optical signals with encoded temporal stimuli. 
We characterise the system's performance 
under single and multiple (wavelength-multiplexed) optical inputs, 
demonstrating resonating spike firing at nanosecond (ns) rates under light-excitation 
in the standard telecom infrared C-band range ($1550$ nm spectral window). 
Furthermore, we utilize the light-sensitive photonic-electronic R\&F RTD neuron of this work 
to realize multiple photon-enabled neuromorphic processing tasks at ns-rates, 
including the chirp test and a temporal pulse separation identification task. 
Moreover, we also demonstrate a digital-to-spike encoding task, 
where a resonate-and-fire light-sensitive neuron 
is combined with time- and wavelength-multiplexing techniques.

\section{Experimental Results}\label{sec2}

The experiments in this work use a photo-detecting RTD 
with a $500$ nm diameter top contact nanopillar  
which incorporates a light absorptive layer 
with a $9$ \textmu m diameter optical window in its structure
(as seen in Fig.\ref{fig:Fig1___Intro}a). 
This absorptive layer, located next to the DBQW, 
is a $250$ nm thick spacer layer (InGaAs)
where most of the electro-hole pair generation occurs.
The DBQW structure is a $5.7$ nm thick quantum well (In$_{0.53}$Ga$_{0.47}$As), 
sandwiched between two $1.7$ nm barriers (AlAs). 
Full details on the RTD and its epi-layer structure can be found in \cite{AlTaai2023}. 
Given the (In,Ga)As-based semiconductor material composition, 
the RTD is permitted to operate at infrared wavelength ranges, 
including the key telecom wavelengths of $1310$ and $1550$ nm \cite{Hejda2022B}. 

Fig.\ref{fig:Fig1___Intro}c 
shows a scanning electron microscope (SEM) image of the device used in this work. 
Fig.\ref{fig:Fig1___Intro}d plots the experimentally measured I-V curves of the RTD 
under dark conditions (blue trace) 
and under illumination of a $0.9$ mW continuous wave (CW) optical signal 
at the infrared wavelength of $1550$ nm (yellow trace). 
These I-V curves shows the existence of a NDR region, 
within which there is a self-oscillatory region
ranging from $V_{RTD} = 0.610$ V to $V_{RTD} = 0.690$ V, 
between the boundaries of the so-called "peak" and "valley" biasing regions. 
Around the self-oscillatory region, 
the system can exhibit deterministic excitable spike firing regimes 
when externally perturbed \cite{Hejda2022B} \cite{Hejda2024_ACSPhotonics}, 
outlining the prospects of RTDs for use as artificial photonic-electronic spiking neurons \cite{Hejda2022A}. 

\subsection{Impulse Test}

\begin{figure}[t!]
\centering
\includegraphics[width=1.00\linewidth]{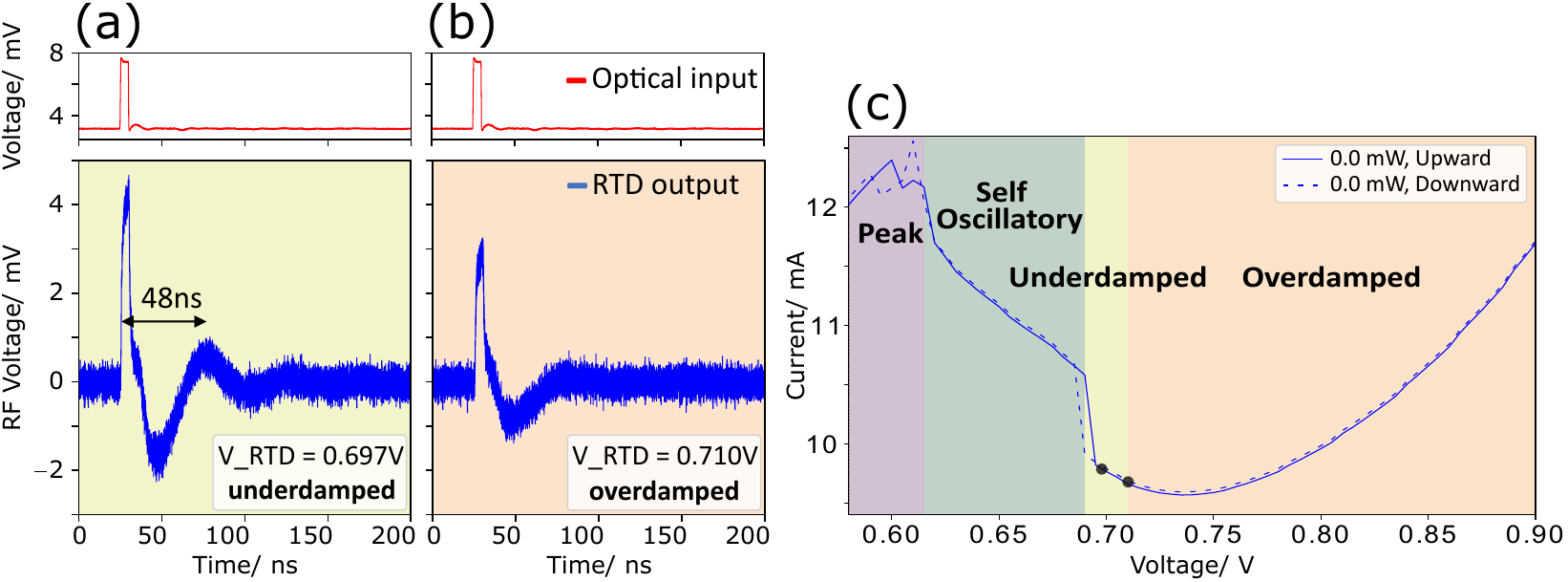}
\caption{\label{fig:Fig2A___ImpTest}
    Impulse response test of the light-sensitive RTD. 
    Sub-threshold optical stimuli are injected into the RTD 
    biased either at (a) $V_{RTD} = 0.697$ V or (b) $V_{RTD} = 0.710$ V in its valley biasing region. 
    Top plots (red time traces) show the optical input signals injected in the RTD, 
    with 5-ns (sub-threshold) square pulses encoded in the optical input 
    (average optical input power of $33.0$ \textmu W). 
    Bottom plots (blue time traces) show the electrical RTD response 
    when biased at $V_{RTD} = 0.697$ V (left) and $V_{RTD} = 0.710$ V (right), 
    revealing the occurrence of underdamped or overdamped oscillations depending on the biasing case. 
    (c) Zoomed in I-V curve of the RTD, 
    marking the two bias voltage points applied to the device for the results in (a) and (b).
}
\end{figure}

\begin{figure}[t!]
\centering
\includegraphics[width=1.0\linewidth]{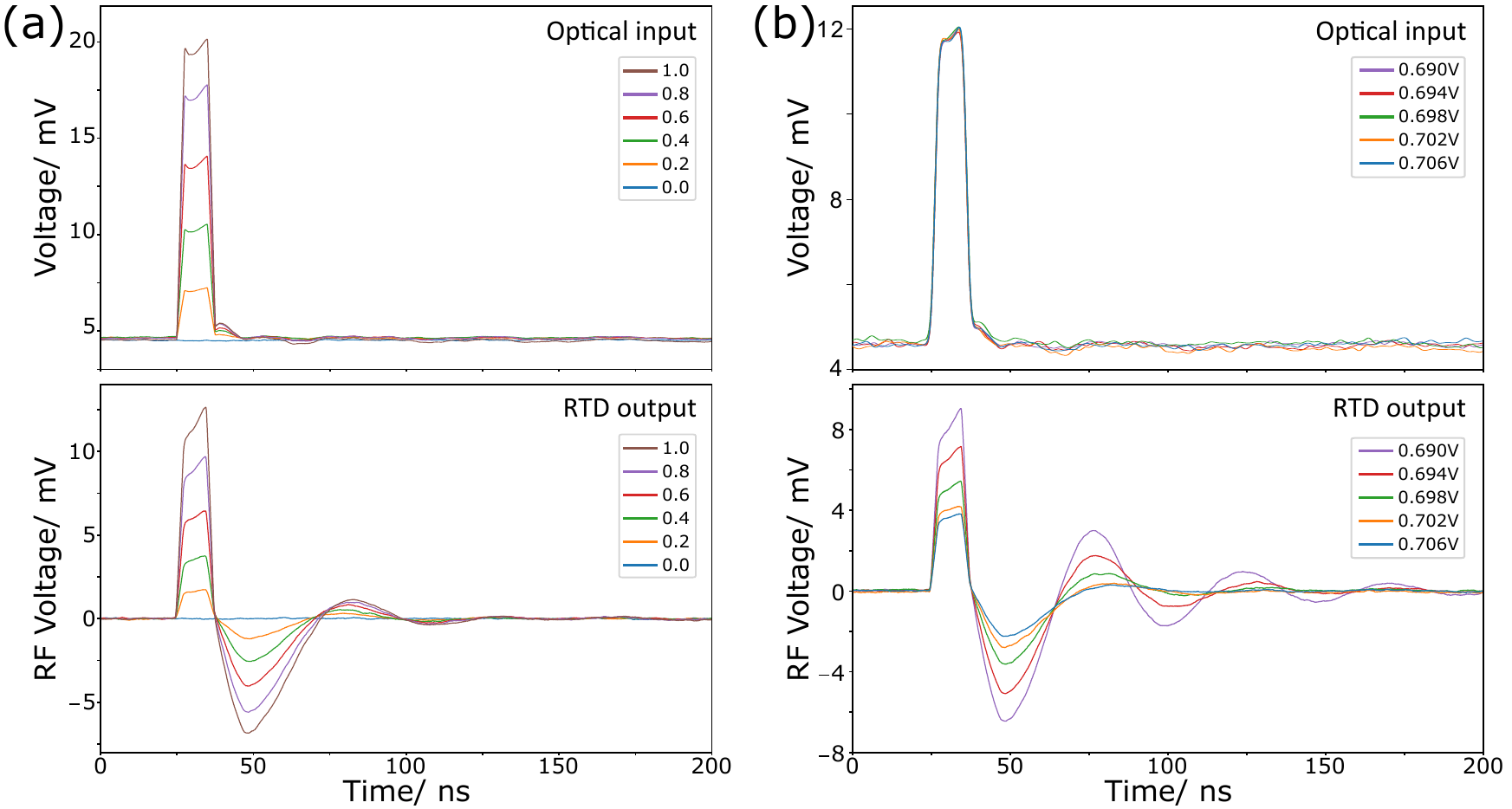}
\caption{\label{fig:Fig2BC___ImptTest}
    Characterisation of the RTD response to injected optical stimuli 
    (average optical input power of $45.0$ \textmu W).
    when reverse biased at the valley operation point.
    a) RTD is biased in the underdamped regime ($V_{RTD} = 0.699$ V) 
    for changing input pulse amplitude.
    b) RTD is injected with input pulses of equal amplitude for changing RTD bias voltage.
}
\end{figure}

To investigate the resonate-and-fire operation of the light-sensitive RTD neuron of this work, 
it is first necessary to perform an impulse test in the system 
(see Figs.\ref{fig:Fig2A___ImpTest}-\ref{fig:Fig2BC___ImptTest}).
The impulse tests permit to analyse the RTD’s temporal output 
when subject to distinct sub-threshold (optical) input perturbations 
and when biased at different voltage points, 
in order to determine the conditions under which resonate-and-fire responses may occur. 
Fig.\ref{fig:Fig2A___ImpTest} shows that the resonance of the RTD circuit can be measured 
by injecting a sub-threshold optical input pulse 
(with insufficient amplitude to trigger a spike firing event in the RTD) 
and observing the voltage decay to the rest state. 
For the test in Fig.\ref{fig:Fig2A___ImpTest}, a signal comprised of 5 ns-long square pulses, 
generated by the arbitrary waveform generator (AWG) in the setup in Fig.\ref{fig:Fig1___Setup}a, 
is encoded onto the light from the tuneable laser using the Mach-Zehnder Modulator (MZM). 
The resulting optical input signal with the encoded sub-threshold optical input pulses 
(see red time-traces in the top panel in Fig.\ref{fig:Fig2A___ImpTest}) 
was injected into the RTD, which was biased in the valley region of its I-V curve 
with two different DC voltages, namely $V_{RTD} = 0.697$ V and $V_{RTD} = 0.710$ V, 
as indicated by black markers in Fig.\ref{fig:Fig2A___ImpTest}c. 
Fig.\ref{fig:Fig2A___ImpTest}a shows that when biased at $V_{RTD} = 0.697$ V, 
the RTD is in the underdamped regime, 
and after the arrival of a sub-threshold optical input pulse, exhibits an oscillatory decay to rest. Fig.\ref{fig:Fig2A___ImpTest}a shows that the resonance period at $V_{RTD} = 0.697$ V bias 
was determined to be 48 ns (peak-to-peak of decay oscillation). 
On the contrary, Fig.\ref{fig:Fig2A___ImpTest}b 
shows that when the RTD is biased at $V_{RTD} = 0.710$ V,
it is said to be in the overdamped regime, 
and after the arrival of the sub-threshold optical input pulse, the decay to rest is exponential. 
The damped oscillatory response highlighted in Fig.\ref{fig:Fig2A___ImpTest} 
underpins the resonate-and-fire functionality in the RTD neuron,  
where spike firing occurs provided that sub-threshold optical inputs 
arrive at specific temporal instants coinciding with the resonance period of the RTD \cite{Hejda2022C}.

Detailed impulse tests, shown in Fig.\ref{fig:Fig2BC___ImptTest}, 
were also performed experimentally on the light-sensitive RTD neuron 
to investigate its reaction to changes 
in both the amplitude of the sub-threshold optical input pulses (Fig.\ref{fig:Fig2BC___ImptTest}a) 
and in the applied bias voltage (Fig.\ref{fig:Fig2BC___ImptTest}b).
First, 10 ns sub-threshold optical input pulses of varying amplitude 
(see top panel in Fig.\ref{fig:Fig2BC___ImptTest}a) were injected into the RTD 
at a fixed bias voltage $V_{RTD} = 0.699$ V 
The response of the system was found to be approximately linear 
where the amplitude of the transient response in the measured output voltage at the RTD 
follows the intensity of the optical input pulses. 
The resonance period of the RTD circuit is found to slightly increase 
(from approx. 55 to 58 ns) as the amplitude of the optical input pulses increases. 
Second, Fig.\ref{fig:Fig2BC___ImptTest}b shows the impulse response of the RTD 
when subject to the injection of 10 ns-long sub-threshold optical pulses of equal amplitude 
(top panel in Fig.\ref{fig:Fig2BC___ImptTest}b) 
into the RTD while the bias voltage is swept from $V_{RTD} = 0.690$ V to $V_{RTD} = 0.708$ V, 
in the so-called valley biasing region. 
The resonance period of the RTD was found to increase with bias voltage (from approx. 52 to 57.5 ns) for the values investigated in Fig.\ref{fig:Fig2BC___ImptTest}b; 
hence, indicating a tuneable resonance frequency in the system. 
Conversely, the RTD response was found to become more damped with higher bias conditions, 
showing decreased oscillation amplitudes. 
Numerical simulations of the impulse response and R\&F spiking regimes in RTD neurons 
have also been performed and are included in the supplementary information accompanying this work
(see Fig.S2). 
The simulations are in good agreement with the results 
observed in the experimental impulse response analysis of the RTD neuron. 
Furthermore, these numerical findings also highlight the potentials of RTD neurons 
for ultrafast operation at sub-ns rates (see Fig.S5).

\subsection{Resonate-and-Fire RTD Neuron}

\begin{figure*}
\centering
\includegraphics[width=0.9\linewidth]{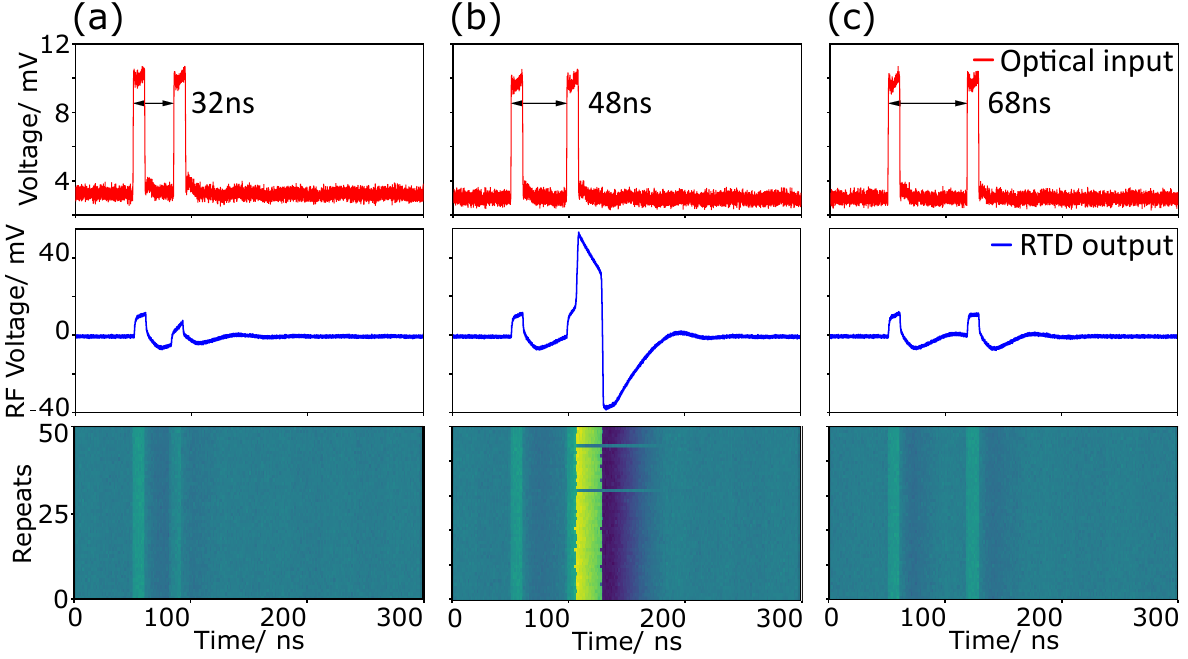}
\caption{\label{fig:Fig3___ResTest}
    Demonstration of the resonate-and-fire effect 
    where the RTD responds to optical stimuli in the form of doublets 
    with different temporal separations when biased in the underdamped regime 
    ($V_{RTD} = 0.699$ V).
    }
\end{figure*}

The impulse test results in Figs.\ref{fig:Fig2A___ImpTest}-\ref{fig:Fig2BC___ImptTest} 
determined that the RTD neuron, when biased in its valley region, 
can indeed undergo resonant responses when stimulated; 
thus, paving the way to demonstrate, for the first time to our knowledge, 
of its operation as a high-speed, efficient, resonate-and-fire photonic-electronic neuron 
for use in neuromorphic computing and sensing functionalities. 
Indeed, the resonate-and-fire effect in the light-sensitive RTD neuron 
is demonstrated experimentally in Fig.\ref{fig:Fig3___ResTest}, 
by observing the interaction of doublet (optical) input stimuli 
and the response provoked at the system’s output. 
To achieve this, pairs of sub-threshold 10 ns-long optical pulses are generated by the AWG 
and encoded onto the intensity of the external tuneable laser. 
These are injected (with average optical power of $50.0$ \textmu W) 
into the RTD (biased at $V_{RTD} = 0.699$ V). 
The temporal separation between the (optical) pulses in the doublet 
is controlled and varied around the value of the inherent resonance of the RTD neuron. 
Here, the temporal separation of the doublet 
is defined as the time between the rising edges of the first and second pulse. 
When the temporal separation between the individual pulses in the doublet 
is less (32 ns, Fig.\ref{fig:Fig3___ResTest}a) or greater (68 ns, Fig.\ref{fig:Fig3___ResTest}c) 
than the resonance period (approx. 48 ns for the RTD circuit of this work, 
see Fig.\ref{fig:Fig2A___ImpTest}a), 
the RTD’s response to the second optical input pulse is decreased 
(as seen in Fig.\ref{fig:Fig3___ResTest}). 
This inhibitory effect occurs since this second optical pulse is incident 
when the RTD is in the antinode (local minima) of the decay oscillation. 
On the other hand, when the separation of the two sub-threshold optical input pulses 
is within the resonance period (48 ns, Fig.\ref{fig:Fig3___ResTest}b), 
the second pulse combines with the decay oscillation (local maxima) arising from the first input pulse. 
This permits the system to cross its activation threshold 
by transiting into its self-oscillatory region, 
triggering therefore an excitable excursion in current-voltage phase space, 
corresponding to firing an electrical spike event as a result of reading voltage with time. 
In this way, the R\&F RTD neuron 
acts like a bandpass filter to incoming (optical) stimuli, 
where spikes are only fired when the frequency of the input (optical) stimuli 
matches the resonance frequency of the RTD neuron. 

\begin{figure}[!t]
\centering
\includegraphics[width=1\linewidth]{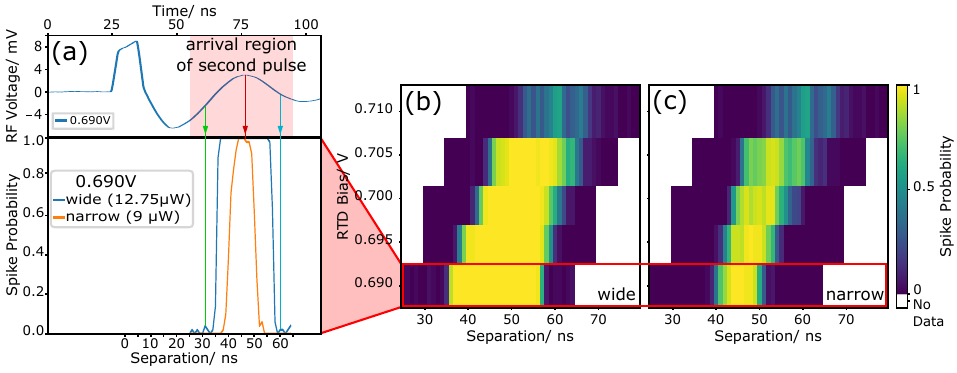}
\caption{\label{fig:Fig4___ResTest-tune}
    Characterisation of the resonate-and-fire effect 
    where the RTD responds to optical stimuli in the form of doublets 
    with different temporal separations for increasing RTD bias and injection power. 
    a) Comparison of the wide and narrow filter for the same bias voltage ($V_{RTD} = 0.690$ V)
    for a tunable bandpass filter width.
    Map of how resonance changes with bias displaying the 
    (b) the widest filter (highest power) and (c) the narrowest filter (lowest usable power).
}
\end{figure}

The experimental results in Figs.\ref{fig:Fig3___ResTest}-\ref{fig:Fig4___ResTest-tune} 
are in excellent agreement with the numerical simulations of the system 
reported in the Supporting Information document (see Fig. S3). 
The theoretical analyses, therefore, validate the system's operation as an R\&F spiking neuron 
exhibiting characteristic frequency-based filtering of input stimuli. 
Furthermore, our simulations also highlight the potentials of system operation 
at even faster timescales (reaching multi-GHz regimes) 
by tuning the RTD's capacitance and inductance values 
(as seen in Fig. S6 in the Supporting Information document). 
These numerical analyses thus highlight the potentials for light-sensitive RTD neurons 
with future circuit device optimisations (e.g. reduced device footprint, circuit parasitics control) 
to act as ultrafast photonic-electronic R\&F neurons 
able to filter optical input stimuli at high-speed information rates.

Upon demonstrating the operation of the RTD as a photonic-electronic R\&F neuron, 
we investigate now the range in which this effect is obtained for different applied bias voltages 
(in its valley biasing region), and highlight its temporal tuneability, 
simply acting on an accessible system parameter (voltage bias). 
To do this, in Fig.\ref{fig:Fig4___ResTest-tune}, 
we measure the RTD’s response to trains of optical input sub-threshold pulse pairs, 
when the device is biased at different voltages. 
Fig.\ref{fig:Fig4___ResTest-tune} shows that when a train of optical input doublets of equal amplitude 
and increasing temporal separation is injected into the RTD, 
the latter will spike only within a determined range of values for pulse pair temporal separation, where the separation coincides with the RTD's damped oscillations frequency range. 
Then, a spike firing probability can be assigned to each temporal separation between pulses, 
as seen in the temporal maps included 
in Fig.\ref{fig:Fig4___ResTest-tune}b and Fig.\ref{fig:Fig4___ResTest-tune}c. 
The centre of the measured resonance window (highlighted in yellow colours in the maps)
corresponds to the maxima of the relaxation oscillation in the RTD, 
and where the achievement of resonate-and-fire response is most probable. 
The resonance window of the RTD neuron increases with optical input power 
since larger pulses can reach the system’s activation threshold 
at lower points on the relaxation oscillation. 

To fully characterise the resonance window of the photonic-electronic RTD neuron at each voltage bias, 
two tests were performed to determine the ‘widest’ and ‘narrowest’ resonance windows. 
First, the so-called ‘widest’ resonance range is measured, 
corresponding to the highest optical injection power set for the optical input pulse trains 
for which the most spikes are triggered from the secondary pulse 
without triggering a spike from the first pulse. 
Next, the so-called ‘narrowest’ resonance window is measured, 
corresponding to the lowest injection power 
for which the least number of consistent spikes are triggered from the secondary pulse. 
Fig.\ref{fig:Fig4___ResTest-tune}a shows the outcome of these two tests 
when the device is biased at the lowest configured voltage of $V_{RTD} = 0.69$ V. 
The top plot in Fig.\ref{fig:Fig4___ResTest-tune}a shows the impulse response of the system, 
whilst the bottom plot shows the obtained ‘widest’ (blue) and ‘narrowest’ (orange) resonant windows 
measured when an average optical input power of $9$ and $12.75$ \textmu W 
was set respectively for the optical input pulse train. 
The latter was formed by optical pulse pairs with temporal separation 
growing from $25$ to $65$ ns  
(red shaded region in the top plot in Fig.\ref{fig:Fig4___ResTest-tune}a) 
around the resonant frequency of the RTD neuron. 
Both the widest and narrowest resonant windows are centred around 46 ns, 
coinciding approximately with the maxima of the resonant oscillation of the system (red arrow). 
For this specific biasing case, 
the approximate ranges for the ‘widest’ and ‘narrowest’ resonant regions 
span from 23 ns to about 11 ns (measured at 0.5 spike firing probability). 
In perfect conditions, the narrowest possible resonance window 
would correspond to the width of the pulse (10 ns in our case), 
with minor differences arising from noise in the system and limitations in the experimental setup. 
The green line and green arrow in Fig.\ref{fig:Fig4___ResTest-tune}a indicate a case 
for which the temporal separation between pulse doublets is below the resonance period of the system. 
In that situation, the second optical input pulse 
will arrive at an instance in time close to the minima in the RTD’s resonant oscillation 
and the system will not elicit a spike firing event (spike probability is zero or close to zero). 
The temporal maps in Fig.\ref{fig:Fig4___ResTest-tune}b and Fig.\ref{fig:Fig4___ResTest-tune}c 
combine in a single plot the responses of the RTD 
when subject to the injection of optical train pulses with increasing temporal separation 
and when the bias voltage is raised from $V_{RTD} = 0.690$ V to $V_{RTD} = 0.710$ V (in steps of 5 mV). 
Fig.\ref{fig:Fig4___ResTest-tune}b and Fig.\ref{fig:Fig4___ResTest-tune}c 
show that the central resonance period of the RTD neuron increases with bias voltage 
from 46 ns (at $V_{RTD} = 0.690$ V) to 54 ns (at $V_{RTD} = 0.710$ V). 
At higher biases, optical injection powers for the ‘widest’ and ‘narrowest’ windows converge 
and spiking becomes less consistent since the oscillations are heavily damped. 
In this way, the resonance window of the RTD neuron acts in its own as a tuneable optical filter, 
whose central resonance frequency and bandwidth can be tuned controllably 
with voltage bias and optical input power, respectively 
(average input optical powers ranging from $9$ to $110$ \textmu W, 
were used across all biases for the results in Fig.\ref{fig:Fig4___ResTest-tune}).

\subsection{Multi-Wavelength Resonate-and-Fire Operation}

\begin{figure}[t!]
\centering
\includegraphics[width=0.75\linewidth]{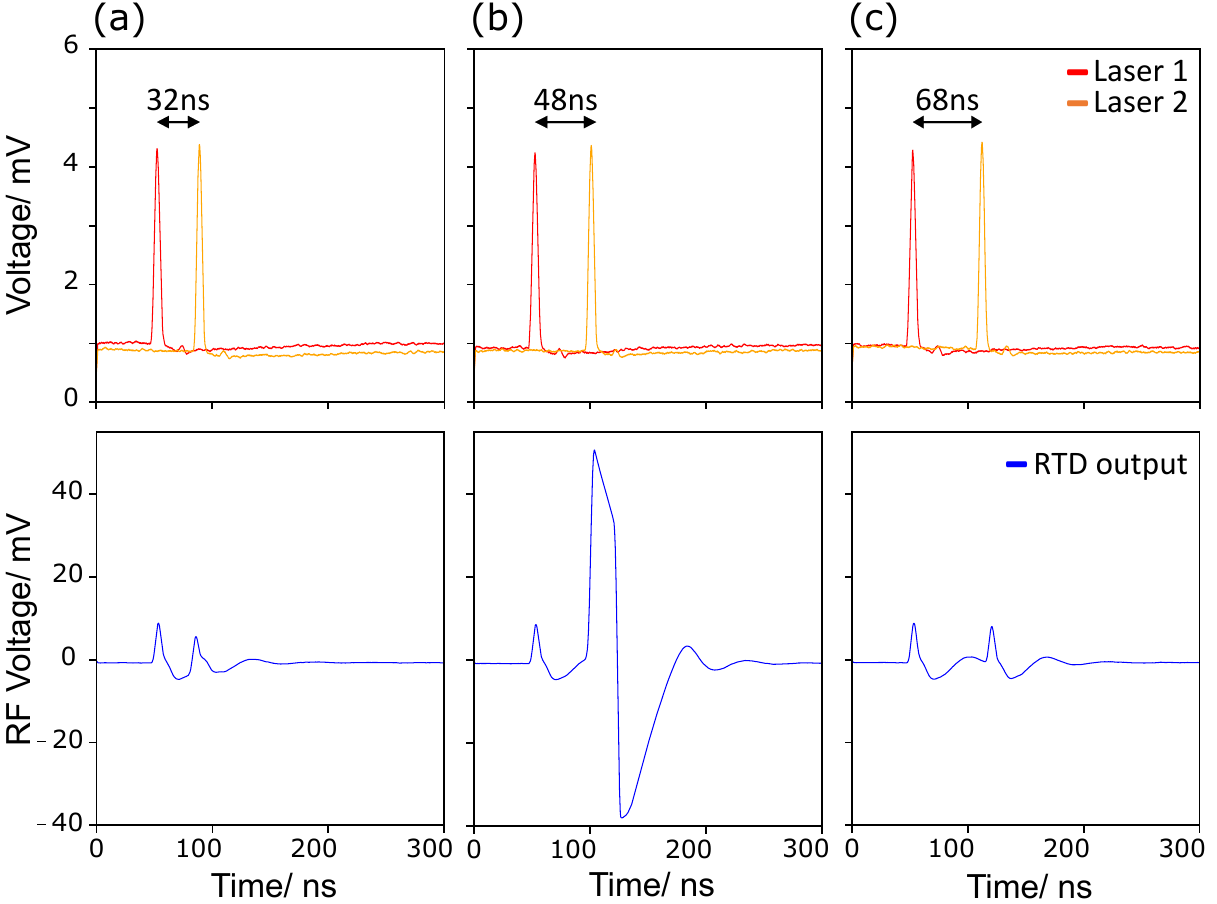}
\caption{\label{fig:Fig7B___SeqTest}
    Demonstration of the resonate-and-fire effect 
    where the RTD responds to optical stimuli from two different VCSELs 
    in the form of 4.8 ns doublets with different temporal separations 
    when biased in the underdamped regime ($V_{RTD} = 0.699$ V).
}
\end{figure}

It is important to note here that the photonic-electronic RTD neuron of this work, 
can be excited by optical input signals across a large wavelength window in the infrared spectrum 
(determined by the bandgap of the light-absorptive layer in its structure) \cite{Romeira2013B}. 
Recent works have demonstrated the possibility to elicit spike firing regimes 
in optically-sensitive (In,Ga)As RTDs
when excited at either $1300$ or $1550$ nm, individually or simultaneously \cite{Hejda2022B}. 
Therefore, given that the system can respond to Wavelength Division Multiplexed (WDM) input signals, 
this allows us to investigate the resonate-and-fire effect in light-sensitive RTD neurons 
subject to a combined optical input signal 
comprised of light from independent sources at different wavelengths. 
For this analysis, we used the experimental setup in Fig.\ref{fig:Fig1___Setup}b 
in which independent VCSELs are used to generate distinct optical input signals 
at various wavelengths 
which are combined into a single line for their injection into the photo-detecting region of the RTD. 
The bias current of each of the VCSELs is direct modulated with RF signals 
to encode square optical input pulses at different time instants 
in different wavelength optical inputs. 
Fig.\ref{fig:Fig7B___SeqTest} shows the experimentally measured response of the RTD neuron 
when subject to WDM-optical pulse doublets, 
where each pulse within the doublet comes from an individual VCSEL at separate wavelength, 
here [$1550$, $1553$] nm. 
The WDM-optical signal is injected into the RTDs photo-absorption layer. 
The average optical power of the combined signal entering the RTD was $23.9$ \textmu W, 
and this was formed by optical pulses set with equal amplitude and growing temporal separation, 
from $32$ ns (Fig.\ref{fig:Fig7B___SeqTest}a), 
to $48$ ns (Fig.\ref{fig:Fig7B___SeqTest}b) and $68$ ns (Fig.\ref{fig:Fig7B___SeqTest}c). 
As seen in Fig.\ref{fig:Fig7B___SeqTest}, 
for this multi-wavelength incoherent optical injection scheme, 
the RTD neuron also fires a spike only when the temporal separation between the doublet optical pulses 
is within the resonance period window of the system (Fig.\ref{fig:Fig7B___SeqTest}b), 
remaining quiescent otherwise 
(Fig.\ref{fig:Fig7B___SeqTest}a and Fig.\ref{fig:Fig7B___SeqTest}c). 
Hence, Fig.\ref{fig:Fig7B___SeqTest} demonstrates 
the multi-wavelength capability of the light-sensitive R\&F RTD neuron of this work; 
thus, permitting to leverage both temporal and wavelength information encoding schemes, 
and therefore enabling increased parallel computational capabilities.

\section{Neuromorphic photonic processing with R\&F RTD neurons}\label{sec3}

Following on the experimental demonstrations of the resonate-and-fire effect 
in light-sensitive RTD neurons, 
both under single (time-multiplexed) and multiple (wavelength-multiplexed) optical input signals 
(see Figs.\ref{fig:Fig3___ResTest}-\ref{fig:Fig7B___SeqTest}), 
we now investigate its application to a variety of light-enabled, neuromorphic/spike-based computational tasks.

\subsection{Chirp Test}

\begin{figure*}[t!]
\centering
\includegraphics[width=1\linewidth]{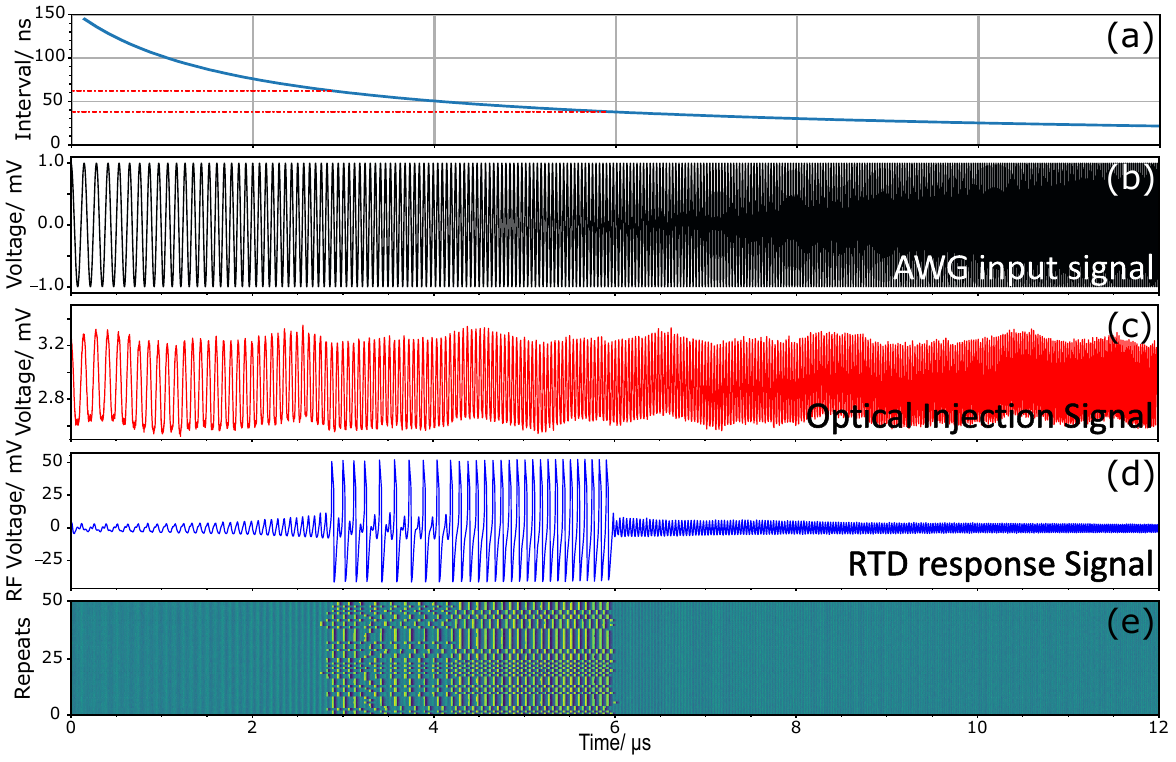}
\caption{\label{fig:Fig5___ChirpTest}
    Chirp Test. 
    (a) Instantaneous period (temporal separation) of the chirp timeseries 
    as a function of time with red dashed lines denoting resonance window. 
    (b) AWG modulation signal for chirp with linearly increasing frequency. 
    (c) Optical injection signal of sinusoid with temporal separation between peaks 
    decreasing inversely with time. 
    (d) RTD response to the injected signal when biased at $V_{RTD} = 0.690$ V. 
    (e) Waterfall plot showcasing the consistency of the measurements across 50 repeats.
}
\end{figure*}

The first task we focused on with the light-sensitive RTD neuron of this work, 
given the bandpass filtering nature of R\&F neurons is the chirp test. 
Chirped signals are utilised in distance-ranging applications, such as SONAR and RADAR \cite{Schiff2000}, 
and in spread spectrum communications 
where a signal is dispersed in the frequency domain of a wider frequency band \cite{Springer2000}. 
This increases the signal range allowing higher resolution detection at greater distances. 

Here, a chirp signal is generated with the instantaneous period of the sinusoid waveform 
decreasing linearly from 150 ns to 25 ns with time, 
over a whole temporal range of $12$ \textmu s, as indicated in Fig.\ref{fig:Fig5___ChirpTest}a. 
The amplitude of the chirp signal remains constant with time. 
Fig.\ref{fig:Fig5___ChirpTest}b plots the chirp signal generated using a 12 GSa/s AWG 
which is encoded optically by modulating (with an MZM) the light from the Tuneable Laser 
in the setup in Fig.\ref{fig:Fig1___Setup}a. 
The time-series in Fig.\ref{fig:Fig5___ChirpTest}c 
plots the features of the optically-encoded chirp signal 
prior to its injection into the R\&F photonic-electronic RTD neuron of this work. 
The latter was biased in its valley region (at a $V_{RTD} = 0.690$ V) 
and the optical signal was set with low average optical power of $95$ \textmu W. 
Fig.\ref{fig:Fig5___ChirpTest}d plots the temporal response of the RTD neuron 
after the injection of the optically-encoded chirp signal. 
Fig.\ref{fig:Fig5___ChirpTest}d reveals that spike firing in the RTD occurs in a well-defined range 
for input waveform periods corresponding to the resonance window of the system (approx. 37-60 ns) 
as seen in Figs.\ref{fig:Fig5___ChirpTest}d and \ref{fig:Fig5___ChirpTest}e. 
Fig.\ref{fig:Fig5___ChirpTest}e plots directly the time-series measured at the output of the RTD, 
whilst Fig.\ref{fig:Fig5___ChirpTest}e provides a temporal map 
gathering in a single plot the measured RTD output for 50 repetitions of the chirp test experiment, 
highlighting the consistency and repeatability of the measured response from the system. 
As seen in Figs.\ref{fig:Fig5___ChirpTest}d and \ref{fig:Fig5___ChirpTest}e, 
the section of the chirp signal with frequencies within the bandpass window of the R\&F RTD neuron 
experiences an increase in oscillation amplitude until the spiking threshold is reached. 
In this way, the RTD neuron is able to filter a stimulus with changing frequency, 
so it only triggers spike firing events for the frequency range 
within the neuron’s resonance frequency window.
This bandpass response to an optical linear chirp signal is also observed 
in the theoretical analyses included in the Supporting Information document (see Fig.S4), 
showing excellent agreement with the experimental findings. 
Moreover, our numerical simulations (see Supporting Information, Fig.S7) 
also predict much faster achievable timescales by varying the RTD's circuit parameters. 
This opens the door for future optimised photonic-electronic RTD systems 
able to operate at multi-GHz rates.

\subsection{Temporal Detection Test}

\begin{figure*}[t!]
\centering
\includegraphics[width=1\linewidth]{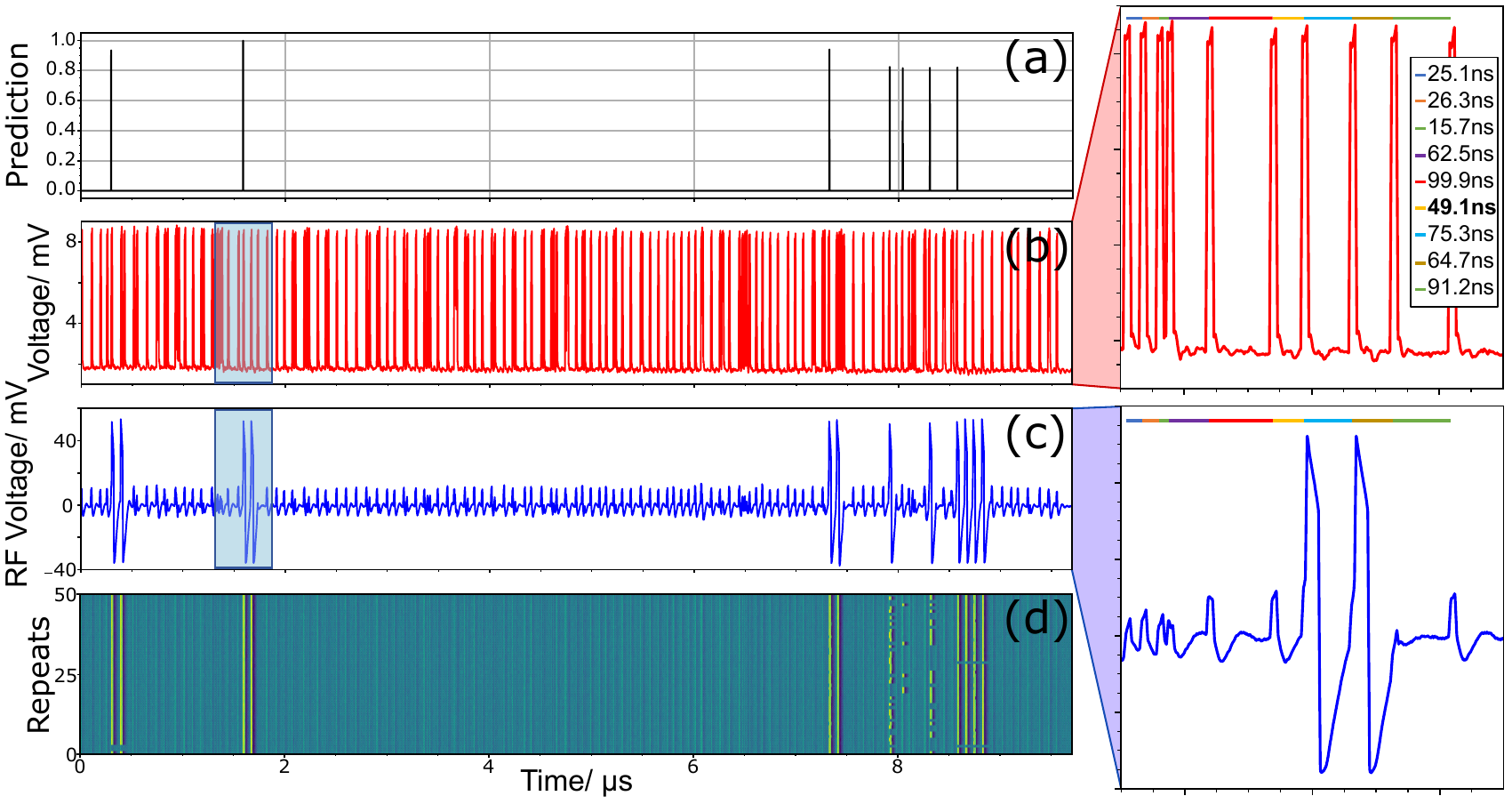}
\caption{\label{fig:Fig6___TempDet}
    Temporal Detection Test. 
    (a) Prediction for spiking based on separation. 
    (b) Optical injection signal of 10 ns pulses with different temporal separations 
    (with zoom of segment). 
    (c) RTD response to the injected signal when biased at $V_{RTD} = 0.699$ V (with zoom of segment). 
    (d) Waterfall plot showcasing the consistency of the measurements across 50 repeats.
    }
\end{figure*}

Following on the chirp test demonstration, the second task we focused on was a temporal detection test, see Fig.\ref{fig:Fig6___TempDet}. 
This consisted of identifying, in a random time-series input signal, 
occurrences of pulse doublets with specific time separations (within the system’s resonance window). 
For this second test, an arbitrary signal formed by a train of pulses 
with random temporal pulse separations ranging between 11-100 ns was generated, 
Fig.\ref{fig:Fig6___TempDet}b. 
This arbitrary signal had a reduced probability of generating pulses doublets 
with temporal separations within the resonance period of the R\&F RTD neuron 
(spanning between 30-60 ns when the device was biased at $V_{RTD} = 0.700$ V), see Fig.\ref{fig:Fig6___TempDet}a. 
The resulting optical input signal in Fig.\ref{fig:Fig6___TempDet}b, 
was generated by externally modulating the light from a tuneable laser source 
for its injection (with an average optical power of $28.3$ \textmu W) 
into the RTD (biased at $V_{RTD} = 0.699$ V). 
Fig.\ref{fig:Fig6___TempDet}c plots the temporal response of the RTD, 
showing the attainment of a spike firing output that acts as a bandpass filter for the optical stimuli, 
as the RTD fires spike events only when the separation between pulses is within its resonance period. 
Further, a clear decreased amplitude RTD response is obtained 
to input pulses with separations smaller or larger than the resonance window of the system, 
as the input pulses arrive before or after the maxima of the relaxation oscillation. 
The time instants where spike firing is achieved (see Fig.\ref{fig:Fig6___TempDet}c), 
coincide with the locations identified in the prediction plot of Fig.\ref{fig:Fig6___TempDet}a, 
which shows the instances where pulses with separations within the resonance period 
are expected to elicit a response. 
Fig.\ref{fig:Fig6___TempDet} therefore demonstrates that the R\&F RTD neuron can indeed 
successfully detect specific temporal features in an optical input time-series signal. 
We must note that, in some cases, a double spike event was obtained 
in response to an input triplet with both a temporal separation within and outwith the resonance period. 
This is because a spike can trigger a subsequent spike 
for a pulse separation outside the resonance window 
since the firing of a spike introduces a phase offset to the relaxation oscillation 
such that the net resonance period immediately after a spike 
(rising edge of spike to first oscillation maxima) 
is greater than the case for after a sub-threshold pulse.
This offset effect caused by spiking can be seen in the inset of Fig.\ref{fig:Fig6___TempDet} 
to be approximately 1.5 times the non-spiking resonance period.
It can also be seen here that instead of firing a third spike, 
the next input pulse arrives early (off resonance) 
and interrupts the relaxation oscillation of the spike.
It should also be noted that the amplitude of the relaxation oscillation after a spike 
is greater than that for after a sub-threshold pulse. 
The temporal map in Fig.\ref{fig:Fig6___TempDet}d merges, in a single plot, 
the measured spiking outputs from the RTD 
when subject to 50 consecutive iterations of the optical input pattern in Fig.\ref{fig:Fig6___TempDet}b. 
The map in Fig.\ref{fig:Fig6___TempDet}d shows the high consistency of the achieved spiking responses 
as highlighted by the patterns (with each spike fired marked by a yellow dot) 
appearing at the desired time locations across all 50 repetitions. 

\subsection{Digital-to-Spike Encoding of Wavelength-Multiplexed Signals}

\begin{figure*}[t!]
\centering
\includegraphics[width=1.00\linewidth]{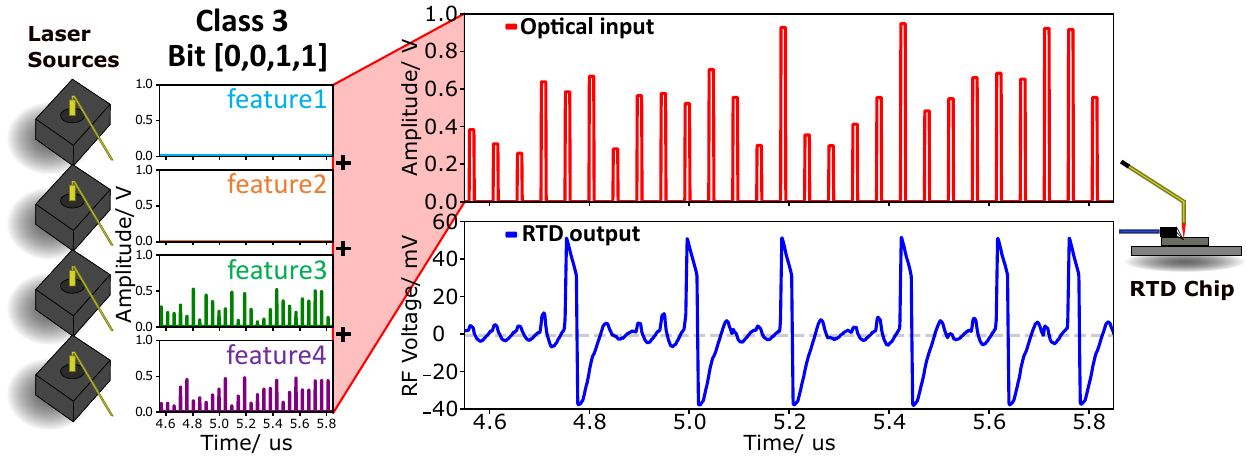}
\caption{\label{fig:Fig8___BitTask1}
    Demonstration of using the resonate-and-fire effect in RTD neurons 
    for spike encoding of a digital signal. 
    (a) The value of each bit corresponds to the modulation amplitude of a mask for each laser. 
    (b) Combined optical injection signal segment of 9.6 ns pulses, 
    separated by the resonance period (48 ns). 
    (c) RTD response signal segment to the injection when biased at $V_{RTD} = 0.696\:V$.
    }
\end{figure*}

We also demonstrate a scheme to encode optical digital input patterns into unique spike patterns 
produced by the resonate-and-fire photonic-electronic RTD neuron. 
Moreover, we utilize the multi-wavelength fan-in operation in the system as the input patterns are delivered 
by different WDM-optical inputs simultaneously injected into the RTD neuron. Specifically, we test the R\&F RTD neuron in the encoding of 4-bit patterns. These are converted into unique temporal spike patterns. 
The latter can subsequently be used to recover the initial 4-bit strings; 
hence, offering a reliable spike encoding-decoding mechanism. 
In this task, each possible combination of 4-bits represents a class 
(such that there are 16 different classes) 
and each of the 4 individual bit values represents a feature 
(such that there are 4 features per class); 
thus, defining a simple 4 feature, 16 class spike encoding task. 
We implement this task using the light-sensitive R\&F RTD spiking neuron of this work, 
when subject to 4 optical input signals generated with 4 separate 1550 nm-VCSELs 
(using the setup in Fig.\ref{fig:Fig1___Setup}b, 
and following the scheme depicted graphically in Fig.\ref{fig:Fig8___BitTask1}). 
The 4 VCSELs are all biased just above their lasing threshold current (I\textsubscript{th}), 
at values less than $1.1\times$I\textsubscript{th} in all cases, 
ensuring all 4 inputs have the same optical power. 
The VCSELs emit all at different wavelengths in the 1550 nm spectral window, 
at [$1550$, $1553$, $1555$, $1558$ nm], respectively. 
The bias current of each VCSEL is direct modulated 
by the means of a 4-channel AWG (Moku:Pro, Liquid Instruments) 
and the four resulting optical inputs are combined into a single line 
for their injection into the light-sensitive RTD neuron.

Fig.\ref{fig:Fig8___BitTask1} depicts the manner in which 4-bit digital optical patterns 
are generated with four independent 1550 nm-VCSELs, 
where each of them carries only the bit values of a given position in the 4-bit strings. 
The features (bit values) are binary, 
and are therefore represented as on-off keying (OOK) of signal from the VCSELs, 
as seen in the example provided in Fig.\ref{fig:Fig8___BitTask1}. 
The latter shows as an example the encoding of the bit string [0,0,1,1], 
also referred here as ‘class 3 pattern’. 
For this specific case, the current modulation applied to the first two VCSELs is ‘off’ 
(thus yielding zero optical power), 
whilst being ‘on’ for the last two (encoding the last two bit positions). 
Each feature (bit value) of the 4-bit string is temporally-masked 
by a set of 128 randomly generated, fixed input weights within the interval [0,1); 
thus, forming a time-multiplexed optical input vector formed by a sequence of 128 virtual nodes, 
with the amplitude of each node directly corresponding to the set input weight value. 
The number of virtual temporal nodes was set to 128 
to fit within the specifications of the memory bank of the 4-channel AWG (Moku:Pro) in the setup. 
The virtual nodes were configured with a total temporal length of 48 ns, 
with a 9.6 ns-long square optical pulse at the onset to encode the input weight value 
and returning to zero for the remainder of the virtual node time. 
This process is depicted graphically in the four inset plots on the left side of Fig.\ref{fig:Fig8___BitTask1}. 
The temporal node length was set to 48 ns to coincide with the resonance period of the RTD neuron, 
whilst also being smaller than its refractory period 
(approx. 100 ns for the device of this work \cite{Hejda2022C}) 
to ensure non-linear temporal coupling between adjacent virtual nodes. 
A total time duration of $6.144$ \textmu s  
was therefore used for the generated 128 time-multiplexed optical input vectors 
representing the individual bits in the 4-bit pattern. 

The waveforms forming the 128 temporal node input vectors encoding the bit values (features) 
were generated by the AWG to directly modulate the bias currents of the VCSELs. 
Each datapoint is normalised such that the resulting combined input is within the range [0,1). 
The optical signals from the four independent 1550 nm-VCSELs in the setup, 
encoding each a different position in the 4-bit string 
(see insets at the left side of Fig.\ref{fig:Fig8___BitTask1}) 
were combined into a single optical line using a four-to-one optical coupler. 
This permitted to directly sum the four weighted features (4-bit values in the string) 
such that each time-multiplexed node receives a contribution from each of the four VCSELs, 
as shown in the top-right plot (red-time series) in Fig.\ref{fig:Fig8___BitTask1}. 
The resulting wavelength-multiplexed optical input signal is then optically injected 
(with average optical power of $24.0$ \textmu W) into the RTD neuron at a rate of 48 ns, 
synchronous to the resonance frequency of the system 
(when biased with a DC voltage of $V_{RTD} = 0.696$ V) 
to ensure the virtual nodes are coupled in time 
through the resonate-and-fire dynamical response in the RTD. 
The optical input power is set to ensure that spikes are triggered in the RTD through the temporal resonate-and-fire effects, rather than just simply thresholding the largest pulses. 

An example of a temporal output response from the RTD-based photonic spiking processor 
is provided in the blue time series in Fig.\ref{fig:Fig8___BitTask1}, 
showing that the system produces a unique temporal electrical spike train pattern 
in response to each specific 4-bit digital optical input signal. 
As the node separation is equivalent to the resonance period, 
a fraction of each node is passed on to the nearest neighbouring node, 
and in that way, short-term memory propagates through the time-connected virtual nodes, 
ensuring nonlinear temporal coupling in the system. 
The nonlinear activation (spiking) function of the R\&F RTD neuron, 
is therefore defined by different time-dependent effects 
(e.g. spike activation, resonance, refractoriness). 
Firstly, this is determined by the intensity coupling between adjacent temporal nodes, 
enabled by relaxation oscillations elicited by sub-threshold optical inputs to the RTD, 
as well as by the relaxation oscillations elicited in the system once a spike has been fired. 
In both cases, these resulting oscillations, of different amplitudes in each case, 
approach or distance the RTD’s operation point from its self-oscillatory region; 
hence modulating effectively the system’s activation threshold 
and modifying dynamically the optical input power level 
required to elicit excitable spike firing events in the RTD. 
Secondly, the inhibitory temporal window defined by the system’s refractory period 
(occurring after a spike is fired) 
also contributes to enhance temporal connectivity between adjacent virtual nodes, 
as in that time, the system remains quiescent 
even when a new super-threshold optical input stimulus enters the system.

\begin{figure*}[t!]
\centering
\includegraphics[width=1.00\linewidth]{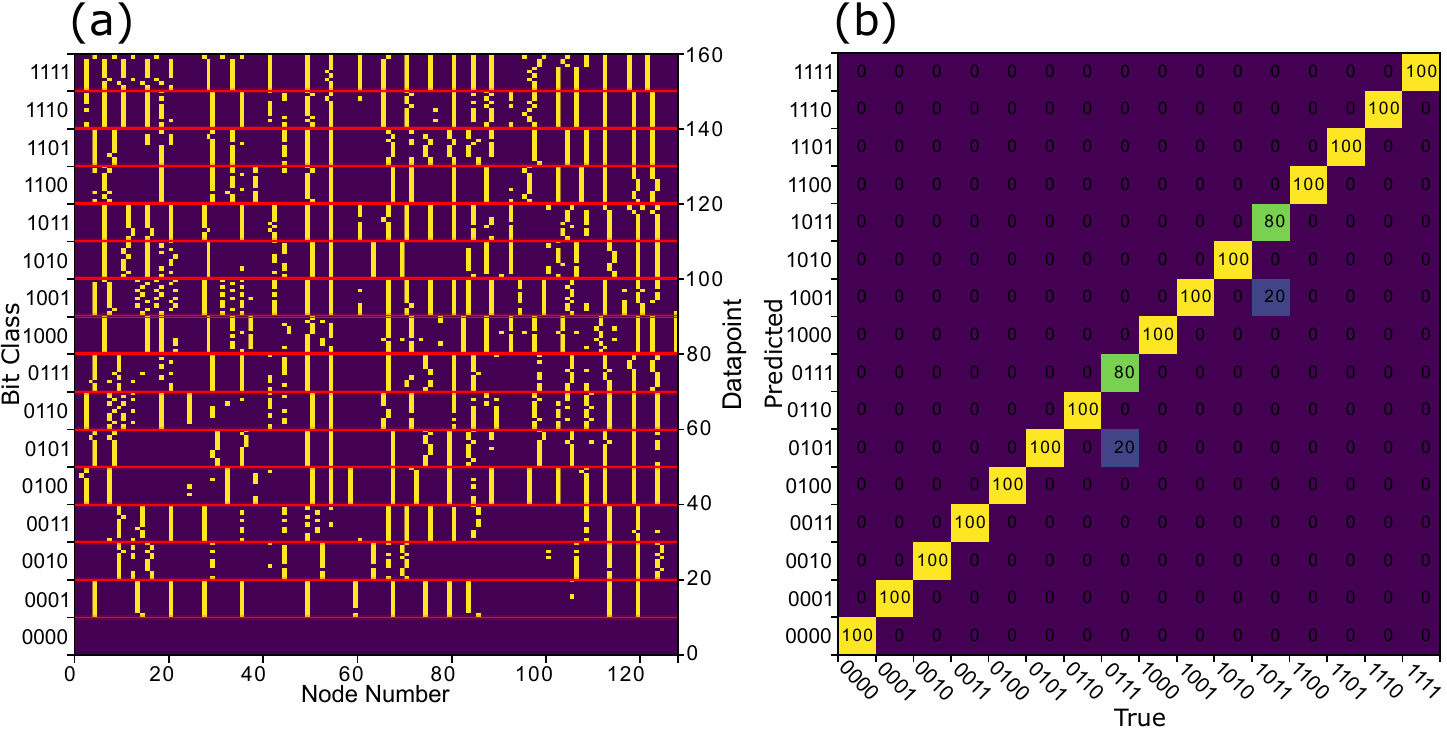}
\caption{\label{fig:Fig9___BitTask2}
    (a) A temporal map of the detected spiking responses from the R\&F RTD neuron 
    for 10 repetitions of each bit string. 
    (b) A typical confusion matrix displaying the trained network's predictions 
    against the true value for each bit string.
    }
\end{figure*}

To demonstrate the operation of the proposed photonic digital-to-spike encoding system, 
ten single (non-time averaged) readings are recorded for each of the 16 classes 
(ranging from ‘0000’ to ‘1111’ and all other cases in between), 
as seen in Fig.\ref{fig:Fig9___BitTask2}. 
These are optically injected into the light-sensitive R\&F RTD neuron 
which produces electrical spike firing temporal outputs in response. 
These are gathered and plotted in the temporal map of Fig.\ref{fig:Fig9___BitTask2}a, 
showing that unique spike patterns are achieved to encode all 16 different classes. 
Additionally, as depicted in Fig.\ref{fig:Fig9___BitTask2}b, 
the temporal spike patterns from the RTD neuron, 
can also be subsequently used to recover the originally-encoded 4-bit patterns. 
This is achieved during a post-processing stage through Linear-least-squares (LLS) fitting 
\cite{Patil2023} \cite{Huang2006}, 
following the method described in \cite{OwenNewns2023A}. 
For this, half of all generated 4-bit spiking patterns are used to train the system, 
and the other half used during the testing stage. 
Fig.\ref{fig:Fig9___BitTask2}b plots the obtained confusion matrix, 
demonstrating successful operation in this 4-bit pattern spike encoding-decoding task, 
yielding an average accuracy of 95\% (98.3\%) with a standard deviation of 2.1\% (1.7\%), 
for train and test stages, respectively, when evaluated over ten data input folds.

\section{Discussion}\label{sec4}

As previously discussed, the resonant tunnelling diode (RTD) is a semiconductor device 
that features a double-barrier quantum well (DBQW) in its structure 
(see diagram in Fig.\ref{fig:Fig1___Intro}b). 
This enables electron quantum tunnelling effects in the system 
that leads to a highly nonlinear current-voltage (I-V) relationship \cite{Ortega2021B}. 
This has a characteristic N-shape, with a region of negative differential resistance (NDR), 
between the so-called ‘peak’ and ‘valley’ local extrema \cite{Ortega2021A}. 
Crucially, there exists a self-oscillatory region within the NDR region, 
around which the system can exhibit neuronal-like dynamical regimes, 
such as excitability and spike firing, 
enabling the use of RTDs as ultrafast artificial neuronal models \cite{Hejda2022A}. 
Notably, RTDs can be designed to be optically sensitive 
by incorporating light-absorption layers in their structure; 
hence, enabling photonic-electronic spiking neuron designs 
that have the highly desirable ability to respond to both photonic and electronic signals
\cite{Romeira2013B}, therefore enabling cascadable interlinking of RTD neurons towards spiking neural networks (SNNs) \cite{Hejda2024_ACSPhotonics}. 

The RTD's ability to exhibit the R\&F effect is mainly due 
to the interplay of the series (quiescent) resistance and the dynamic resistance.  
In the self-oscillation section of the NDR region, 
the negative dynamic resistance dominates the positive series resistance, 
hence the I-V gradient (differential resistance) is negative. 
In the underdamped region, the negative dynamic resistance is no longer sufficient
to compensate for the positive series resistance, 
so damped oscillations are observed. 
As the DC bias is moved farther away from the NDR, 
this positive series resistance increases and the negative dynamic resistance decreases, 
leading to a reduction in oscillation amplitude. 
The valley point is reached when the negative dynamic resistance becomes zero.
This resonant behaviour is only observed in certain kind of RTDs 
when reverse biased at the valley operation point.
This effect is not observed in the peak region (before reaching the NDR region)
since both resistances are positive (there is only addition). 

The underlying mechanism for R\&F behaviour relates to the Andronov-Hopf bifurcation type. 
In the RTD, this bifurcation presents as a bistable region in the NDR of the N-shaped I-V curve.
Here, this bistable region, located between the peak/valley voltage points, 
is a consequence of the asymmetric shape of the NDR \cite{Ortega2021A}.
For the RTD in Fig.\ref{fig:Fig1___Intro}d, 
the bistable region appears particularly pronounced in the valley region). 
In this bistable region, both a stable fixed point and a stable limit cycle coexist.
Evidently, there also exists an unstable limit cycle between these 
that acts as a threshold for the system entering the stable limit cycle.
Note that, perturbing the system into this stable limit cycle trajectory 
is equivalent to triggering a spike.
Consequently, after a subthreshold perturbation, 
the system relaxes toward its fixed point while oscillating a few times around it, 
which enables operation in the R\&F mode. 

Importantly, Fig.\ref{fig:Fig1___Intro}d shows that external light injection into the RTD 
affects the I-V curve (shifting it upwards and also slightly laterally).
Here, electron-hole pair generation increases the current in the device, 
which reduces its series resistance, leading to this voltage and current shift.
This allows for the system to also be triggered with external optical perturbations 
to directly prompt light-elicited spike firing regimes in RTD neurons \cite{Figueiredo1999}.



\section{Conclusion}\label{sec5}

This work demonstrates experimentally, for the first time, 
a photonic-electronic resonate-and-fire spiking neuron. 
This is realized using an optically-sensitive resonant tunnelling diode (RTD) 
allowing operation at key infrared telecom wavelengths in the C-band (1550 nm). 

This RTD neuron displays all-or-nothing spiking responses at nanosecond rates, 
and experiences an oscillatory decay to steady state when externally perturbed (optically) 
in the so-called valley region of its highly nonlinear, N-shaped I-V characteristic. 
Theoretical analyses are also carried out in this work, showing excellent agreement with the experimental findings.
The excitability resonance period of the RTD neuron, 
ultimately determined by the RCL parameters of the circuit, 
is observed to be on the order of tens-of-nanoseconds for our current experimental setup. 
However, potentially much faster (sub-ns) rates are viable, as demonstrated numerically in this work, through circuit parameter and circuit parasitics optimisation. 
Moreover, we also found that both the resonance period 
and the damping of the relaxation oscillations in the RTD neuron 
can be tuned by acting on the applied bias voltage. 
This provides a basis for a frequency-sensitive event-based spike encoding mechanism, 
where the precise timing of input stimuli is of significant importance and unlocks extra neuron capabilities not found in the LIF neuron model.

For the photonic-electronic R\&F neuron of this work, 
deterministic firing of electrical spikes from pairs of sub-threshold optical pulses is demonstrated, where a spike is fired only when the inter-pulse time interval of the stimuli 
falls within the narrow resonance window of the neuron. 
The bandpass filtering abilities of the neuron were experimentally characterised 
for injection power and bias, with resonance period increasing with bias, 
and the resonance window increasing with injection power. 
Furthermore, the multi-wavelength operation of the RTD neuron was demonstrated. 
The wide absorption band of the inbuilt photosensitive epi-layer in the device
allows for spike triggering by multiple modulated optical input signals of different wavelengths.

Finally, multiple practical tasks are experimentally demonstrated 
at fast nanosecond rates for the photonic-electronic R\&F RTD neuron. 
These include a spike-based bandpass filtering of optically-encoded chirp signals, 
as well as in a temporal detection test, 
where spiking is observed only in the sections of the optical input signal 
matching the RTD neuron excitability resonance window. 
A proof-of-concept demonstration for a neuromorphic photonic-electronic spike encoding scheme 
is also shown, 
where masked optical input bit-strings are transformed into event-based electrical spike patterns, 
utilizing the temporally-dependent, nonlinear and stateful (short-term memory) character 
of responses of the resonate-and-fire effect. 
The spike-encoded sequence patterns can be trained 
to recover the original bit-string with accuracy 98.3\%.

\section{Experimental Methods}\label{sec6}

\begin{figure*}[t!]
\centering
\includegraphics[width=1.0\linewidth]{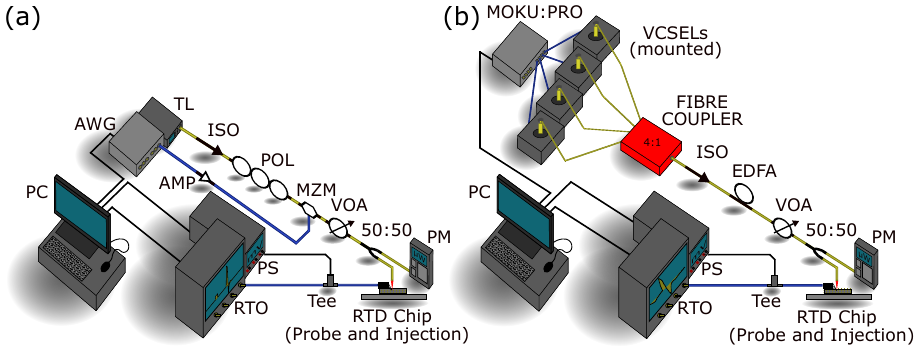} 
\caption{\label{fig:Fig1___Setup}
    Experimental setups used to investigate the R\&F RTD neuron of this work 
    under optical injection of fast intensity encoded stimuli from 
    (a) an externally-modulated 1550 nm tuneable laser and 
    (b) four directly-modulated 1550 nm-Vertical-Cavity Surface Emitting Lasers (VCSELs) 
    multiplexed in wavelength.
    Arbitrary waveform generator (AWG), four-channel AWG (MOKU:PRO), RF amplifier (AMP), 
    tuneable laser (TL), optical isolator (ISO), polarisation controller (POL), 
    Mach-Zehnder Modulator (MZM), variable optical attenuator (VOA), 
    fibre-optic splitter (50:50), four-to-one fibre-optic coupler (4:1),
    Erbium-Doped Fibre Amplifier (EDFA), power supply (PS), bias-tee (TEE), 
    power-meter (PM), real-time oscilloscope (RTO), computer (PC).
    }
\end{figure*}

In this work, we demonstrate, for the first time, the operation of an RTD 
as a photonic resonate-and-fire spiking neuron. 
Moreover, we demonstrate this functionality 
when the system is subject to single or multiple optical inputs at high speeds. 
The externally optical injected signals are generated using either 
an externally-modulated tuneable laser 
or four directly-modulated Vertical-Cavity Surface Emitting Lasers (VCSELs), 
shown in Fig.\ref{fig:Fig1___Setup}a and Fig.\ref{fig:Fig1___Setup}b, respectively. 

Firstly, we investigate the photonic R\&F neuron capabilities of the RTD
when subject to a single ns-duration optical input signal. 
To this end, we used the setup in Fig.\ref{fig:Fig1___Setup}a 
in which the CW output from a 1550 nm tuneable laser (TL) 
is externally modulated using a Mach-Zehnder Modulator (MZM). 
The electrical RF signals used to modulate the TL’s light 
are generated by a high-speed (12 GSa/s, 5GHz bandwidth) arbitrary waveform generator (AWG), 
then sent to an RF amplifier (AMP) prior to their application to the MZM. 
The optical isolator (ISO) in the setup reduces unwanted back-reflections 
and the polarisation controller (POL) matches the polarisation of the TL’s light to that of the MZM 
to minimise power losses in the system. 
A variable optical attenuator (VOA) is used 
to control the injection power of the optical input signal. 
The optical path is then divided via a 50:50 fibre-optic splitter, 
with the first branch connected to a power-meter (PM) or a photodetector (PD) 
to measure the average power or monitor the optical injection signal, respectively. 
The second branch is optically injected into the light-sensitive RTD 
by means of a lensed optical fibre. 
The RTD is biased with a DC voltage generated from a power supply (PS) 
and a bias-tee (TEE) is used to allow the temporal responses of the RTD neuron 
to be read-out with a 16 GHz real-time oscilloscope 
for their subsequent capture and analysis with a computer (PC).

Additionally, in this work we also investigated the operation of the R\&F RTD neuron 
when subjected simultaneously to multiple optical inputs at different wavelengths. 
For this purpose, we used the experimental setup included in Fig.\ref{fig:Fig1___Setup}b, 
in which four telecom (1550 nm) VCSELs generate four independent optical input signals. 
These are biased just above their lasing threshold current 
(in the range of 2 mA at the temperature of $25\:^\circ$C for all VCSELs)
with values of [2.26, 2.02, 2.50, 2.27] mA, respectively. 
The VCSELs showed lasing emission 
at the wavelengths of [1550, 1553, 1555, 1558] nm, respectively, 
mimicking a Wavelength Division Multiplexing (WDM) system with approx. $330$ GHz channel spacing. 
The current of each VCSEL was directly modulated using separate RF signals generated 
with a four channel 1.25 GSa/s AWG (Moku:Pro, Liquid Instruments). 
A four-to-one fibre-optic coupler is used to merge, in a single optical path, 
the light emission from the four individual VCSELs, 
creating effectively a WDM injection scheme. 
This strategy permits a multi-wavelength fan-in functionality in the RTD 
for the generated optical signals from the VCSELs. 
An Erbium-Doped Fibre Amplifier (EDFA) is utilised 
to boost the combined WDM optical input power from the four VCSELs 
and to overcome losses in the experimental setup. 

\section{Acknowledgements}\label{sec7}

The authors acknowledge support from 
the UKRI Turing AI Acceleration Fellowships Programme (EP/V025198/1), 
the UKRI 'ProSensing' project (EP/Y030176/1), 
the UK Multidisciplinary Centre for Neuromorphic Computing (UKRI982), 
the EU Pathfinder Open project ‘SpikePro’ (Grant ID 101129904), 
from the Funda\c{c}\~ao para a Ci\^encia e a Tecnologia 
(FCT, projects 2022.03392.PTDC – META-LED, and ERC-PT-A), 
and from the Fraunhofer Centre for Applied Photonics, FCAP.


\bibliography{sn-bibliography}

\end{document}